\documentstyle[12pt]{article}
\begin{document}

\begin{flushright}
NDA-FP-21 \\
OCHA-PP-65 \\
August 1995 \\
hep-th/9508112
\end{flushright}

\vfill

\begin{center}
{\bf Collective Motion of Micro-organisms \\
from Field Theoretical Viewpoint}

\vfill

{\sc Shin'ichi NOJIRI}\footnote{
E-mail: nojiri@cc.nda.ac.jp}

{\it Department of Mathematics and Physics,
National Defence Academy \\
Yokosuka, 239 JAPAN}

and

{\sc Masako KAWAMURA}\footnote{
E-mail: masako@phys.ocha.ac.jp} and
{\sc Akio SUGAMOTO}\footnote{E-mail:
sugamoto@phys.ocha.ac.jp}

{\it Department of Physics, Ochanomizu University \\
1-1, Otsuka 2, Bunkyo-ku, Tokyo, 112, JAPAN}

\vfill

{\bf ABSTRACT}

\end{center}

We analyze the collective motion
of micro-organisms in the fluid and
consider the problem of the
red tide. The red tide is produced by
the condensation of the
micro-organisms, which might be a similar
phenomenon to the condensation of
the strings. We propose a model of the generation of the red tide.
By considering the interaction between the micro-organisms mediated
by the velocity fields in the fluid,  we derive the Van der Waals
type equation of state, where the generation of the red tide can be
regarded as a phase transition from the gas of micro-organisms
to the liquid.

\newpage

String theory is a candidate of the unified theory
of elementary particles and their interactions.
If we can derive from the string theory at the Planck
scale the field contents and the mass ratios
at TeV scale,
it might be possible to confirm that nature is really
described by the string theory.
In order to get a stronger confidence, however,
we need to know the dynamics of the string theory,
{\it e.g.}, the mechanisms of the compactification,
supersymmetry breaking {\it etc.}

In this respect it is important to study a
system which has some of
the properties similar to string theories.
The techniques obtained from string theories might
be applied to analyze such a system and some implications
to string theory might be derived from the analysis
of the system.
One of such a systems is the micro-organisms swimming in the
fluid.
The problem of the swimming of the micro-organisms was
considered from the gauge theoretical viewpoint by Shapere
and Wilczek \cite{sw1}\cite{sw2} and from the string and
membrane theoretical viewpoint in Refs.\cite{ksn} and \cite{nks}.

In this paper, we consider the problem of the red tide.
The red tide is generated by the condensation of the
micro-organisms, which might be a similar phenomenon
to the condensation of the strings.
Having such a view, we propose a model of the red tide,
where we take into account the
interaction between the micro-organisms mediated by the
velocity fields in the fluid and derive the Van der Waals
type equation.
In this model, the generation of the red tide is regarded as
a phase transition from the gas of micro-organisms
to the liquid.

The Reynolds number is very small for the micro-organisms
swimming in the fluid and the Navier-Stokes equation
is linearized.
Then the equations of motion for an incompressible fluid
are given by
\begin{eqnarray}
\label{incom}
\nabla\cdot{\bf v}&=&0 \ , \\
\label{press}
\Delta{\bf v}&=&{1 \over \eta}\nabla p\ \ {\rm or}\ \
\Delta(\nabla\times{\bf v})=0\ .
\end{eqnarray}
Here ${\bf v}$, $p$ and $\eta$ are the velocity field,
the pressure and the coefficient of viscosity, respectively.
In two dimensions, these equations (\ref{incom}) and
(\ref{press}) have the following forms:
\begin{eqnarray}
\label{incom2}
\partial_z v_{\bar z}
+\partial_{\bar z}v_z&=&0 \ , \\
\label{press2}
4\eta\partial_z\partial_{\bar z}v_{\bar z}=
\partial_{\bar z} p \ , \ &{\rm and}&\
4\eta\partial_z\partial_{\bar z}v_z =
\partial_z p \ .
\end{eqnarray}
It is an interesting point that the
equations of motion (\ref{incom2}) and (\ref{press2}) can
be derived from the QED-like Lagrangean in the Landau gauge:
\begin{equation}
\label{lag}
{\cal L}=2\eta
(\partial_z v_{\bar z}-\partial_{\bar z}v_z)^2
-p(\partial_z v_{\bar z}
+\partial_{\bar z}v_z)\ .
\end{equation}
Furthermore, the Lagrangean (\ref{lag})
can be regarded as the entropy density of the fluid,
that will be shown next.

Following the standard textbook of the fluid
dynamics,\footnote{
We follow the notation of Landau-Lifsitz's one}
we find the time-derivative of the fluid entropy $S$ is given by
\begin{equation}
\label{entropy}
\dot S=\int dV\Bigl\{ {\sigma'_{ik} \over 2kT}\Bigl(
{\partial v_i \over \partial x_k}
+{\partial v_k \over \partial x_i}\Bigr)
-{{\bf q}\cdot {\rm grad}(kT) \over (kT)^2}\Bigr\}\ .
\end{equation}
Here $\sigma'_{ik}$ is the stress
tensor (in three dimensions):
\begin{equation}
\label{stress}
\sigma'_{ik}=\eta\Bigl(
{\partial v_i \over \partial x_k}
+{\partial v_k \over \partial x_i}
-{2 \over 3}\delta_{ik}
{\partial v_l \over \partial x_l}\Bigr)
+\zeta\delta_{ik}
{\partial v_l \over \partial x_l}\ .
\end{equation}
If we assume that the temperature $T$ is globally
defined; ${\rm grad}(kT)=0$ and the fluid is incompressible
(\ref{incom}), the equation (\ref{entropy}) can be rewritten
as
\begin{equation}
\label{entropy2}
\dot S=\int dV{\eta \over 2kT}\Bigl\{ \Bigl(
{\partial v_i \over \partial x_k}
-{\partial v_k \over \partial x_i}\Bigr)^2
+p{\partial v_l \over \partial x_l}
+{\partial \over \partial x_i}\Bigl(
v_k{\partial v_i \over \partial x_k}\Bigr)\Bigr\}\ .
\end{equation}
Here we have imposed the condition of the
incompressibility by the multiplier field $p$.
The resulting expression is equivalent to the action
(\ref{lag}) of
QED in the Landau gauge if we neglect the surface term.

The micro-organism appears as a boundary condition.
\begin{equation}
\label{boundary}
{\bf v}({\bf x}={\bf X}(t;{\bf \xi}))
=\dot {\bf X}(t;{\bf \xi}) \ .
\end{equation}
Here the shape of the surface of a micro-organism is
described by a mapping function; ${\bf X}(t;{\bf \xi})$.
Here ${\bf \xi}$ parametrizes the surface of the
micro-organism. The equation (\ref{boundary}) tells that
there is no slipping between the surface of a
micro-organism and the sticky fluid.
In the following, we only consider the two dimensional case
for simplicity. The extension to the three dimensional one
is tedious but straightforward.
Let a micro-organism exist at the origin $z=0$. Then by solving
the equations (\ref{incom2}), (\ref{press2}) under
the boundary condition (\ref{boundary}), we obtain
\begin{equation}
v_{\bar z}=\phi_1(z)-z\overline{\phi_1'(z)}+\overline{\phi_2(z)}\ .
\end{equation}
Here $\phi_i$'s are analytic functions:
\begin{equation}
\label{mode}
\phi_1(z)=\sum_{k<0}a_kz^{k+1}\ , \hskip 1cm
\phi_2(z)=\sum_{k<-1}b_kz^{k+1}\ .
\end{equation}
The coefficients $a_k$ and $b_k$ can be determined by the
boundary condition (\ref{boundary}).
By evaluating the singularity of the velocity fields
at the origin, we obtain
\begin{eqnarray}
\label{multi}
&& -4\partial_z\partial_{\bar z}v_{\bar z}
+{1 \over \mu}\partial_{\bar z}p \nonumber \\
&& =4\pi\Bigl\{\sum_{k<-2}{2(-1)^k a_k \over (-k-2)!}
\delta^{(-k-1)}(\bar z)\delta (z) \nonumber \\
&& +\sum_{k \leq -3}{\bar b_k(-1)^k \over (-k-2)!}
\delta^{(-k-1)}(\bar z)\delta (z)
-\sum_{k \leq -3}{b_k(-1)^k \over (-k-2)!}
\delta (\bar z) \delta^{(-k-1)}(z) \Bigr\} \ .
\end{eqnarray}
Note that the l.h.s in Eq.(\ref{multi}) is given
from the Lagrangean (\ref{lag}) by the variation
with respect of the velocity fields. Therefore
the r.h.s. represents the source terms and we find
that the micro-organisms can be expressed as
multi-pole source.

When there are two micro-organisms, we can evaluate
the interaction action $S^{\rm int}$
between them by using Eq.(\ref{entropy2});
\begin{eqnarray}
\label{int}
S^{\rm int}&=&-{16\eta \over kT}
\int dt \sum_{k,l\leq -2}
\pi(k+1)(l+1){(-1)^l (-l-k-2)! \over (-k-1)!(-l-1)!} \nonumber \\
&& \hskip 1cm \times (a_k^{(1)}a_l^{(2)}z_0^{k+l+1}\bar z_0
+\bar a_k^{(1)}\bar a_l^{(2)}z_0 \bar z_0^{k+l+1})\ .
\end{eqnarray}
Here $i$ in $a_n^{(i)}$ and $b_n^{(i)}$ expresses
$i$-th ($i=1,2$) micro-organism,
$z_0$ is the relative coordinate
between the two micro-organisms and $k$ is the Boltzman constant.

We now consider a model of the generation of the red tide.
For simplicity, we assume $N$ micro-organisms with radius
$r_0$ and mass $m$ move by using a single mode $a_l$
($l\neq -1$) in Eq.(\ref{mode}).
Then the micro-organism can rotate but cannot translate
by themselves. Then the partition function of the system is
given by
\begin{equation}
\label{partition}
Z={1 \over N!}
\int_0^\infty {\rm e}^{-{1 \over kT}\sum_{i=1}^N
{p_i^2 \over 2m}}
\Omega^{{\rm fluid}}dz_1d\bar z_1\cdots dp_{z_N}dp_{\bar z_N}\ .
\end{equation}
Here $\Omega^{{\rm fluid}}={\rm e}^{- S^{{\rm int}}}$
can be evaluated by using Eq.(\ref{int}),
\begin{eqnarray}
\label{int2}
S^{{\rm int}}&=& \sum_{i,j}{\phi(
{\bf r}_{ij}) \over kT}
\nonumber \\
&=& -{16\eta \over kT}\int dt \sum_{i,j}
\pi (l+1)^2{(-1)^l (-2l-2)! \over {(-l-1)!}^2} \nonumber \\
&& \hskip 1cm \times
(a_l^{(i)}a_l^{(j)}z_{ij}^{2l+1}\bar z_{ij}
+\bar a_k^{(i)}\bar a_l^{(j)}z_{ij} \bar z_{ij}^{2l+1})\ .
\end{eqnarray}
Here ${\bf r}_{ij}=(z_{ij}, \bar z_{ij})$ is the relative coordinate
between the $i$-th micro-organism and $j$-th one.
After the momentum integration, we obtain the following expression
\begin{equation}
\label{part2}
Z=(2\pi mkT)^N \Omega(N,T,V)\ .
\end{equation}
Here $V$ is the volume of the system.
If we assume the interaction is small,
${\rm e}^{-\phi({\bf r}_{ij})}-1\ll 1$,
$\Omega(N,T,V)$ can be approximated by the two body interaction;
\begin{equation}
\label{part3}
\Omega(N,T,V)={V^N \over N!}\Bigl\{1+{N^2 \over 2V}\int_0^{2\pi}d\theta
\int_0^\infty dr\, r\Bigl({\rm e}^{-
{\phi({\bf r}_{ij}) \over kT}}-1
\Bigr)+\cdots \Bigr\}\ .
\end{equation}
Here $z_{ij}=r{\rm e}^{i\theta}$. By integrating over
the relative angle
$\theta$ between two micro-organisms, the second term in
Eq.(\ref{part3}) has the following form;
\begin{equation}
\label{part4}
\int_0^{2\pi}d\theta\int_0^\infty dr\, r\Bigl({\rm e}^{-{\phi({\bf r}_{ij})
\over kT}}-1
\Bigr)=2\pi\int_0^\infty dr\, r \Bigl\{
I_0\Bigl({C_l \over kT}r^{2l+2}\Bigr)-1\Bigr\}\ .
\end{equation}
Here $C_l$ is a positive quantity defined by
\begin{equation}
\label{cl}
C_l\equiv 32\eta \pi (l+1)^2
{(-2l-2)! \over \{(-l-1)!\}^2 } \int dt |a_l^{(i)}(t)a_l^{(j)}(t)|
\
\end{equation}
and $I_0(x)\equiv \int_0^{2\pi}{d\theta \over 2\pi}{\rm e}^{x\cos\theta}$
is a deformed Bessel function.
Since $I_0(x)$ is a monotonically increasing function,
Eq.(\ref{part4}) tells
that the interaction between two micro-organisms becomes effectively
attractive one after being averaged with respect to $\theta$.

Since the micro-organism has a finite radius $r_0$, the potential energy
becomes effectively $+\infty$ inside the radius $r_0$; $r<r_0$. Therefore
the equation (\ref{part4}) should be modified by
\begin{eqnarray}
\label{part5}
&& \int_0^{2\pi}d\theta\int_0^\infty dr\, r\Bigl(
{\rm e}^{-{\phi({\bf r}_{ij}) \over kT}}-1
\Bigr) \nonumber \\
&& \hskip 1cm =2\pi\int_0^{r_0} dr\, r (-1)+ 2\pi\int_{r_0}^\infty dr\, r
\Bigl\{
I_0\Bigl({C_l \over kT}r^{2l+2}\Bigr)-1\Bigr\}\ .
\end{eqnarray}
Furthermore if we consider the case that ${C_l \over kT}r_0^{2l+2}\ll 1$,
we can approximate the deformed Bessel function
$I_0\Bigl({C_l \over kT}r^{2l+2}\Bigr)
\sim 1+{1 \over 4}\Bigl({C_l \over kT}r^{2l+2}\Bigr)^2$ and we obtain
\begin{equation}
\label{part6}
\int_0^{2\pi}d\theta\int_0^\infty dr\, r\Bigl({\rm e}^{-{\phi({\bf r}_{ij})
\over kT}}-1
\Bigr)=
{2 \over N}\Bigl\{ b-{a \over (RT)^2}\Bigr\}\ .
\end{equation}
Here $a$ and $b$ are positive quantities defined by
\begin{eqnarray}
\label{ab}
a&\equiv& -{\pi N^3 \over 2(2l+3)}\Bigl( {C_l \over 2}
\Bigr)^2r_0^{4l+6} \ ,\\
b&\equiv& {N \over 2}\pi r_0^2 \ ,
\end{eqnarray}
and $R=Nk$
The $b$ is nothing but the volume occupied by all ($N$)
the micro-organisms.
Since the free energy $F$ and the pressure (of micro-organism)
$p$ are defined by
\begin{eqnarray}
F&\equiv&-Nk\ln Z \ , \\
p&\equiv&-\Bigl({\partial F \over \partial V} \Bigr)_T \ ,
\end{eqnarray}
we obtain
\begin{equation}
\label{eqstate}
p+{a \over V^2RT}={RT \over V}\Bigl(1 + {b \over V}\Bigr)
\end{equation}
We can replace the factor $\Bigl(1 + {b \over V}\Bigr)$
in the r.h.s. of Eq.(\ref{eqstate}) by
$\Bigl(1 - {b \over V}\Bigr)^{-1}$
since the effect of $b$ can be considered to be the
decrease of the volume $V$ of the system by the occupation
of the micro-organisms.
Then we obtain the following Van der Waals type equation
of state for the micro-organism gas;
\begin{equation}
\label{eqstate2}
p={RT \over V-b}-{a \over V^2 RT}\ .
\end{equation}
In the usual Van der Waals equation, the second term
in Eq.(\ref{eqstate2}) is not $-{a \over V^2 RT}$ but
$-{a \over V^2}$.
The qualitative feature of the equation obtained here is,
however, similar to the usual Van der Waals equation.
If we consider $p$-$V$ curve with fixed $T$,
there appears a region where
${\partial p \over \partial V}>0$ when
$T<T_c={2 \over 3R}\sqrt{a \over 3b}$.
The appearance of such an unphysical region implies
the existence of the phase transition between liquid and gas.
The phase transition would  be interpreted as the generation of the
red tide.
For fixed $T<T_c$, the phase transition from gas to liquid
occurs when the volume decreases, which would correspond to
the increase of the number density of the micro-organisms.
Furthermore $T_c$ becomes higher as $a$ becomes bigger.
Since $a$ is proportional to
$|a_l|^4$, the red tide might be generated if the micro-organisms
become to move more actively when the temperature of the
water rises
higher in summer.

If we assume that Eq.(\ref{eqstate2}) is valid even in
three dimensional space, we can estimate the order of the
number density of the micro-organisms when the phase
transition occurs. Eq.(\ref{eqstate2}) tells that the
phase transition occurs when
\begin{equation}
\label{temp}
T\sim {1 \over R}\sqrt{{a \over V}}\ .
\end{equation}
By letting $l$ and $t$ be the typical length and time
scales of micro-organisms and by using the dimensional
analysis, we obtain
\begin{equation}
\label{a}
a\sim N^3\eta^2 l^9 t^{-2}\ .
\end{equation}
By substituting Eq.(\ref{a}) to Eq.(\ref{temp}),
we find
\begin{equation}
\label{temp2}
T\sim {\eta \over k}l^{{9 \over 2}}t^{-1}\sqrt{n}\ .
\end{equation}
Here $n={N \over V}$ is the number density of the
micro-organisms.
Since the Boltzman constant $k$ and the coefficient
of viscosity $\eta$ for water is given by
\begin{eqnarray}
&& k\sim 10^{-23}{\rm JK^{-1}}\sim 10^{-16}
{\rm erg K^{-1}} \\
&& \eta\sim 10^{-2}{\rm g\, cm^{-1}s^{-1} }\ ,
\end{eqnarray}
the number density $n$ is given by
\begin{equation}
\label{ndens}
n\sim 10^{-6} {\rm cm^{-3}}\ .
\end{equation}
Here we have assumed $l\sim 10^{-2}{\rm cm}$ and
$T\sim 10^2{\rm K}$.
Eq.(\ref{ndens}) tells that the phase transition occurs
when there are several micro-organisms in 1 ${\rm m}^3$.
Since we have neglected several numerical factors, some of
which depend on the details of the motion of
the micro-organisms, the obtained number density would not
be so unreasonable.

\end{document}